\documentclass[aps,twocolumn,showpacs]{revtex4}
\usepackage{amsmath}
\usepackage{graphicx}
\usepackage{dcolumn}
\usepackage{verbatim}
\usepackage{times}
\usepackage{subfigure}
\usepackage{bm}
\usepackage{color}
\usepackage[colorlinks,dvipdfm]{hyperref}
\usepackage{subfigure}

\begin{document}

\title{D-wave superconductivity induced by short-range antiferromagnetic
correlations in the two-dimensional Kondo lattice model}
\author{Yu Liu$^{1}$, Huan Li$^{1}$, Guang-Ming Zhang$^{1}$, and Lu Yu$^{2}$}
\affiliation{$^{1}$State Key Laboratory of Low-Dimensioanl Quantum Physics and Department
of Physics, Tsinghua University, Beijing 100084, China;\\
$^{2}$Institute of Physics and Institute of Theoretical Physics, Chinese
Academy of Sciences, Beijing 100190, China.}
\date{\today }

\begin{abstract}
The possible heavy fermion superconductivity is carefully
reexamined in the two-dimensional Kondo lattice model with an
antiferromagnetic Heisenberg superexchange between local magnetic
moments. In order to establish an effective mean field theory in
the limit of the paramagnetic heavy Fermi liquid and near the
half-filling case, we find that the spinon singlet pairing from
the local antiferromagnetic short-range correlations can reduce
the ground state energy substantially. In the presence of the
Kondo screening effect, the Cooper pairs between the conduction
electrons is induced. Depending on the ratio of the Heisenberg and
the Kondo exchange couplings, the resulting superconducting state
is characterized by either a d-wave nodal or d-wave nodeless
state, and a continuous phase transition exists between these two
states. These results are related to some quasi-two dimensional
heary fermion superconductors.
\end{abstract}

\pacs{71.27.+a, 74.70.Tx, 75.30.Mb}
\maketitle

\section{Introduction}

Heavy fermion materials have been playing a particular important role in our
understanding of strongly correlated electron systems\cite{Stewart,Lohneysen}%
, and the Kondo lattice model is believed to capture the basic physics of
heavy fermion systems\cite{Si-2001}. The model describes a lattice of local
spin-1/2 magnetic moments coupled antiferromagnetically to a single band of
conduction electrons. When the number of conduction electrons is less than
the number of the local magnetic moments, the coherent superposition of
individual Kondo screening clouds gives rise to the huge mass enhancement of
the quasiparticles, and the resulting metallic state is characterized by a
\textit{large} Fermi surface with the Luttinger volume containing both
conduction electrons and local moments\cite{Ogata-2007,Assaad-2008}.
Competing with the Kondo singlet formation, the local magnetic moments
indirectly interact with each other via magnetic polarization of the
conduction electrons -- the Ruderman-Kittel-Kasuya-Yosida (RKKY)
interaction. Such an interaction dominates at low values of the Kondo
exchange coupling and is the driving force for the antiferromagnetic (AFM)
long-range order near the half-filling of the conduction electrons \cite%
{Doniach,Lacroix-Cyrot,zhang-2000}.

In addition, there have been growing evidence that the antiferromagnetism is
also intimate with superconductivity in some typical heavy fermion
compounds, such as CeCu$_{2}$Si$_{2}$ (Ref.\cite{Steglich-1979}) and CeRhIn$%
_{5}$ (Ref.\cite{Parks-2006}). So far various mechanisms for heavy fermion
superconductivity have been studied, including paramagnon exchange\cite%
{Anderson,Scalapino}, conventional phonon-mediated\cite{Fulde},
and Kondo-boson-mediated
pairings\cite{Auerbach-1986,Millis-1986,Lavagna}. So far the
pairing mechanism of the heavy quasiparticles is still under
investigation theoretically. However, various experimental
evidences in some heavy fermion materials strongly support the
d-wave pairing superconductivity\cite{Thalmeier-Steglich}.

It is well established that the large-$N$ fermionic approach can be used to
treat the Kondo lattice model very efficiently, leading to a stable
paramagnetic heavy Fermi liquid state\cite%
{read-newns,Millis-1986,Auerbach-1986}. Qualitatively, the RKKY interaction
promotes singlet formation among the local magnetic moments, reducing the
tendency of singlet formation between the local moments and conduction
electrons. To simplify the RKKY interaction, one can explicitly introduce
the local AFM Heisenberg superexchange $J_{H}$ among the local moments to
the Kondo lattice system\cite%
{Coleman-1989,Coqblin-1997,Senthil-2004,Coleman-2005,Pepin-2008,Senthil-2009,zhang-2011}%
. Thus, the paramagnetic heavy Fermi liquid state ($J_{K}>J_{H}$) provides a
good starting point for theoretical considerations, because a number of
instabilities can be further discussed, including the AFM ordered state and
unconventional heavy fermion superconductivity. Recently, some numerical
evidences on robust d-wave pairing correlations have been given in small
size Kondo-Heisenberg lattice system\cite{Dagotto}.

In this paper, we try to establish an effective mean field (MF) theory to
the Kondo-Heisenberg lattice model on a two-dimensional square lattice in
the limit of $J_{K}\gg J_{H}$. After carefully examining three different MF
treatments for the local AFM Heisenberg superexchange, we find that the
spinon or f-fermion pairings can substantially save the ground state energy
in the presence of the Kondo screening. Near the half-filling, the Cooper
pairs between the conduction electrons can be induced via the Kondo
screening effects. We further find that the resulting superconducting
pairing function has a d-wave symmetry. Whether there are nodes or not
depends on the ratio of the local AFM Heisenberg and the Kondo exchange
couplings $x=J_{H}/J_{K}$. There is a continuous phase transition between
the nodal and nodeless superconducting phases. Before the magnetic phase
transition, the Kondo screening order parameter is strongly suppressed, and
the present mean field theory is no longer valid.

\section{Mean field treatment}

The model Hamiltonian of the Kondo-Heisenberg lattice model is given by:
\begin{equation}
H=\sum_{\mathbf{k},\sigma }\epsilon _{\mathbf{k}}c_{\mathbf{k}\sigma
}^{\dagger }c_{\mathbf{k}\sigma }+J_{K}\sum_{i}\mathbf{S}_{i}\cdot \mathbf{s}%
_{i}+J_{H}\sum_{\left\langle ij\right\rangle }\mathbf{S}_{i}\cdot \mathbf{S}%
_{j},
\end{equation}%
where $c_{k\sigma }^{\dagger }$ creates a conduction electron on an extended
orbital with wave vector $\mathbf{k}$ and z-component of spin $\sigma
=\uparrow ,\downarrow $. The spin-1/2 operators of the local magnetic
moments have the fermionic representation $\mathbf{S}_{i}=\frac{1}{2}%
\sum_{\sigma \sigma ^{\prime }}f_{i\sigma }^{\dagger }\mathbf{\tau
}_{\sigma \sigma ^{\prime }}f_{i\sigma ^{\prime }}$ where
$\mathbf{\tau }$ is the Pauli matrices. There are two local
constraints: $\sum_{\sigma }f_{i\sigma }^{\dagger }f_{i\sigma }=1$
and $f_{i\uparrow }f_{i\downarrow }=0$. The former local
constraint restricts any charge fluctuations, the f-fermions just
describe the spin degrees of freedom of the local moments, and we
will refer them to spinons. The latter constraint is imposed by
the spin SU(2) symmetry and the extended s-wave spinon pairing
between spinons can be excluded. So the paramagnetic Fermi liquid
limit ($J_{K}\gg J_{H}$) provides us a good starting point of
theoretical considerations.

It is first noticed that the Kondo spin exchange can be expressed as the
singlet pairing attraction between the spinon holes and conduction electrons
up to a chemical potential shift
\begin{equation}
\mathbf{S}_{i}\cdot \mathbf{s}_{i}=-\frac{1}{2}\left( f_{i\uparrow
}^{\dagger }c_{i\uparrow }+f_{i\downarrow }^{\dagger }c_{i\downarrow
}\right) \left( c_{i\uparrow }^{\dagger }f_{i\uparrow }+c_{i\downarrow
}^{\dagger }f_{i\downarrow }\right) .
\end{equation}%
Then the Kondo screening order parameters can be introduced as
\begin{equation}
V=\left\langle f_{i\uparrow }^{\dagger }c_{i\uparrow }+f_{i\downarrow
}^{\dagger }c_{i\downarrow }\right\rangle .
\end{equation}%
For the non-magnetic states, the local AFM Heisenberg superexchange can be
expressed in terms of either the spinon hopping or singlet pairing operators
\begin{align}
\mathbf{S}_{i}\cdot \mathbf{S}_{j}& =-\frac{1}{2}(f_{i\uparrow }^{\dagger
}f_{j\uparrow }+f_{i\downarrow }^{\dagger }f_{j\downarrow })(f_{j\uparrow
}^{\dagger }f_{i\uparrow }+f_{j\downarrow }^{\dagger }f_{i\downarrow })+%
\frac{1}{4}  \notag \\
& =-\frac{1}{2}(f_{i{\uparrow }}^{\dag }f_{j\downarrow }^{\dag
}-f_{i\downarrow }^{\dag }f_{j{\uparrow }}^{\dag })({{f}_{j\downarrow }{f}_{i%
{\uparrow }}-{f}_{j{\uparrow }}{f}_{i\downarrow })+}\frac{1}{4}.
\end{align}%
However, most of previous theoretical studies on the Kondo-Heisenberg
lattice model\cite%
{Coqblin-1997,Senthil-2004,Coleman-2005,Pepin-2008,Senthil-2009,zhang-2011},
only introduced the spinon hopping order parameter. Actually,
Coleman and Andrei\cite{Coleman-1989} had emphasized that the
local SU(2) gauge invariance of the local Heisenberg spin operator
generally requires the consideration of both MF order parameters.
Moreover, recent advances in this area have been made by using\
symplectic representation of the local magnetic spins
\cite{Coleman-2010}. Here we first introduce two MF valence bond
order parameters to characterize the short-range AFM correlations
between the local moments%
\begin{equation}
\chi _{ij}=-\langle f_{i\uparrow }^{\dagger }f_{j\uparrow }+f_{i\downarrow
}^{\dagger }f_{j\downarrow }\rangle ,\Delta _{ij}=-\langle f_{i{\uparrow }%
}^{\dag }f_{j\downarrow }^{\dag }-f_{i\downarrow }^{\dag }f_{j{\uparrow }%
}^{\dag }\rangle .
\end{equation}

To avoid the incidental degeneracy of the conduction electron band on a
square lattice, we choose $\epsilon _{\mathbf{k}}=-2t\left( \cos k_{x}+\cos
k_{y}\right) +4t^{\prime }\cos k_{x}\cos k_{y}-\mu $, where $t$ and $%
t^{\prime }$ are the first and second nearest neighbor hoping matrix
elements, while $\mu $ is the chemical potential, which should be determined
self-consistently by the density of conduction electrons $n_{c}$. When the
local AFM correlation hopping parameter $\chi $ is simply chosen as a
uniform parameter, the spinons form a very narrow band with a dispersion $%
\chi _{\mathbf{k}}=J_{H}\chi (\cos k_{x}+\cos k_{y})+\lambda $, where $%
\lambda $ is the Lagrangian multiplier to impose the local constraint $%
\sum_{\sigma }f_{i\sigma }^{\dagger }f_{i\sigma }=1$ on average.
For the short-range AFM spinon singlet pairing order parameter
$\Delta _{ij}$, the local pairing constraint $f_{i\uparrow
}f_{i\downarrow }=0$ excludes the extended s-wave pairing, and
then d-wave symmetric pairings lead to a lower ground state
energy\cite{Kotliar-Liu}, corresponding to $\Delta
_{i,i+e_{x}}=-\Delta _{i,i+e_{y}}\equiv \Delta _{0}$.

Then the MF model Hamiltonian in momentum space can be written as
\begin{align}
H{_{mf}}& =\sum_{\mathbf{k}\sigma }\varepsilon _{\mathbf{k}}c_{\mathbf{k}%
\sigma }^{\dag }c_{\mathbf{k}\sigma }-\frac{J_{K}V}{2}\sum_{\mathbf{k}\sigma
}\left( f_{\mathbf{k}\sigma }^{\dag }c_{\mathbf{k}\sigma }+c_{\mathbf{k}%
\sigma }^{\dag }f_{\mathbf{k}\sigma }\right)  \notag \\
& \text{ \ \ }+J_{H}\sum_{\mathbf{k}}\Delta (\mathbf{k})(f_{\mathbf{k}{%
\uparrow }}^{\dag }f_{-\mathbf{k}\downarrow }^{\dag }+h.c.)+N\epsilon _{0}
\notag \\
& \text{ \ \ }+\sum_{\mathbf{k}\sigma }\chi _{\mathbf{k}}f_{\mathbf{k}\sigma
}^{\dag }f_{\mathbf{k}\sigma },
\end{align}%
where $\Delta _{\mathbf{k}}\equiv \Delta _{\mathbf{0}}(\cos k_{x}-\cos
k_{y}) $ and $\epsilon _{0}=(\frac{J_{K}V^{2}}{2}+J_{H}\Delta
_{0}^{2}+J_{H}\chi ^{2}-\lambda )$. When a Nambu spinor is defined by $\psi
_{\mathbf{k}}^{\dagger }=\left(
\begin{array}{cccc}
c_{\mathbf{k}\uparrow }^{\dag } & c_{-\mathbf{k}\downarrow } & f_{\mathbf{k}%
\uparrow }^{\dag } & f_{-\mathbf{k}\downarrow }%
\end{array}%
\right) $, we can express the MF model Hamiltonian in a matrix form%
\begin{align}
H{_{mf}}& =\sum_{\mathbf{k}}\psi _{\mathbf{k}}^{\dagger }\left(
\begin{array}{cccc}
\varepsilon _{\mathbf{k}} & 0 & -\frac{J_{K}V}{2} & 0 \\
0 & -\varepsilon _{\mathbf{k}} & 0 & \frac{J_{K}V}{2} \\
-\frac{J_{K}V}{2} & 0 & \chi _{\mathbf{k}} & J_{H}\Delta _{\mathbf{k}} \\
0 & \frac{J_{K}V}{2} & J_{H}\Delta _{\mathbf{k}} & -\chi _{\mathbf{k}}%
\end{array}%
\right) \psi _{\mathbf{k}}  \notag \\
& \text{ \ \ \ \ }+\sum_{\mathbf{k}}\left( \varepsilon _{\mathbf{k}}+\chi _{%
\mathbf{k}}\right) +N\epsilon _{0},
\end{align}%
Diagonalizing this MF model Hamiltonian, we obtain two quasiparticle energy
bands%
\begin{equation}
E_{\mathbf{k}}^{\pm }\equiv \sqrt{E_{\mathbf{k}1}\pm \sqrt{E_{\mathbf{k}%
1}^{2}-E_{\mathbf{k}2}^{2}}},
\end{equation}%
where%
\begin{align}
E_{\mathbf{k}1}& \equiv \frac{1}{2}\left( \varepsilon _{\mathbf{k}}^{2}+\chi
_{\mathbf{k}}^{2}+J_{H}^{2}\Delta _{\mathbf{k}}^{2}\right) +\frac{1}{4}%
(J_{K}V)^{2},  \notag \\
E_{\mathbf{k}2}& \equiv \sqrt{(\varepsilon _{\mathbf{k}}\chi _{\mathbf{k}}-%
\frac{J_{K}^{2}V^{2}}{4})^{2}+(\varepsilon _{\mathbf{k}}J_{H}\Delta _{%
\mathbf{k}})^{2}}.  \label{spectrum}
\end{align}%
Due to the particle-hole symmetry of the superconducting quasiparticles, all
the negative energy states are filled up at zero temperature, and the ground
state energy per site is thus given by
\begin{equation}
E_{g}=\frac{1}{N}\sum_{\mathbf{k}}\left( \varepsilon _{\mathbf{k}}-E_{%
\mathbf{k}}\right) +\frac{J_{K}V^{2}}{2}+J_{H}\left( \Delta _{0}^{2}+\chi
^{2}\right) ,
\end{equation}%
where $E_{\mathbf{k}}\equiv E_{\mathbf{k}}^{+}+E_{\mathbf{k}}^{-}=\sqrt{2E_{%
\mathbf{k}1}+2E_{\mathbf{k}2}}$. The saddle point equations for the MF order
parameters $V$, $\chi $, $\Delta _{0}$ and $\lambda $ can be determined by
minimizing the ground state energy:
\begin{equation}
\frac{\partial E_{g}}{\partial V}=\frac{\partial E_{g}}{\partial \chi }=%
\frac{\partial E_{g}}{\partial \lambda }=\frac{\partial E_{g}}{\partial
\Delta _{0}}=0.
\end{equation}%
The chemical potential $\mu _{c}$ is determined by the relation $n_{c}=-%
\frac{\partial E_{g}}{\partial \mu _{c}}$. Therefore, the self-consistent
equations at zero temperature are given by,
\begin{widetext}
\begin{align}
\frac{1}{N}\sum_{\mathbf{k}}\frac{1}{E_{\mathbf{k}}}\left[ \varepsilon _{%
\mathbf{k}}+\frac{\chi _{\mathbf{k}}(\varepsilon _{\mathbf{k}}\chi _{\mathbf{%
k}}-J_{K}^{2}V^{2}/4)+\varepsilon _{\mathbf{k}}J_{H}^{2}\Delta _{\mathbf{k}%
}^{2}}{E_{\mathbf{k}2}}\right] & =\left( 1-n_{c}\right) ,  \notag \\
\frac{1}{N}\sum_{\mathbf{k}}\frac{1}{E_{\mathbf{k}}}\left[ \chi _{\mathbf{k}%
}+\frac{\varepsilon _{\mathbf{k}}(\varepsilon _{\mathbf{k}}\chi _{\mathbf{k}%
}-J_{K}^{2}V^{2}/4)}{E_{\mathbf{k}2}}\right] & =0,  \notag \\
\frac{1}{N}\sum_{\mathbf{k}}\frac{1}{E_{\mathbf{k}}}\left[ 1-\frac{%
(\varepsilon _{\mathbf{k}}\chi _{\mathbf{k}}-J_{K}^{2}V^{2}/4)}{E_{\mathbf{k}%
2}}\right] & =\frac{2}{J_{K}},  \notag \\
\frac{1}{N}\sum_{\mathbf{k}}\frac{\Delta _{\mathbf{k}}^{2}}{E_{\mathbf{k}}}%
\left[ 1+\frac{\varepsilon _{\mathbf{k}}^{2}}{E_{\mathbf{k}2}}\right] & =%
\frac{2\Delta _{0}^{2}}{J_{H}},  \notag \\
\frac{1}{N}\sum_{\mathbf{k}}\frac{\left( \chi _{\mathbf{k}}-\lambda \right)
}{E_{\mathbf{k}}}\left[ \chi _{\mathbf{k}}+\frac{\varepsilon _{\mathbf{k}%
}(\varepsilon _{\mathbf{k}}\chi _{\mathbf{k}}-J_{K}^{2}V^{2}/4)}{E_{\mathbf{k%
}2}}\right] & =2J_{H}\chi ^{2}.
\end{align}
\end{widetext}

In order to simplify the present treatments, we can have three different MF
schemes: (1) $\chi =0$ and $\Delta _{0}\neq 0$; (2) $\chi \neq 0$ and $%
\Delta _{0}\neq 0$; (3) $\chi \neq 0$ and $\Delta _{0}=0$. Without losing
generality, we choose $t^{\prime }/t=0.3$, $n_{c}=0.8$ and fixing $%
J_{K}/t=2.0$, we have carefully solved the self-consistent equations and
compared the ground state energy densities for the three different types of
MF schemes. The numerical results are not sensitive to the parameters
chosen. The numerical results are displayed in Fig. 1. It is clear that, for
$0<J_{H}/t<1.3$, the first type of MF ansatz has a comparatively lower
ground state energy. As the local AFM coupling $J_{H}/t$ increases, the
ground state energy is almost unchanges, and then decreases gradually as $%
J_{H}/t$ is further increased. However, the ground state energies for two
other MF schemes grow up in the beginning and then blend down after reaching
their maximal points. Only when $J_{H}/t>1.34$, the ground state energy of
the second MF ansatz starts to become lower than that of the first MF
ansatz. In the present paper, we will confine our following discussions to
the small local AFM Heisenberg exchange coupling $J_{H}/t<1.34$, so only the
first type of MF scheme ($\chi =0$ and $\Delta _{0}\neq 0$) will be used.

\begin{figure}[tbp]
\includegraphics[scale=0.4]{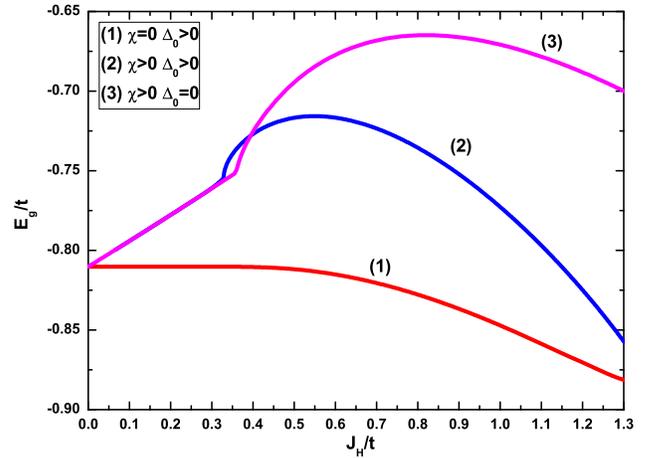}
\caption{(Color online) The ground state energies of a two-dimensional
Kondo-Heisenberg lattice model under three different types of MF
approximation schemes for a given value of $J_{K}/t=2.0$. The other
parameters are $t^{\prime }/t=0.3$ and $n_{c}=0.8$.}
\end{figure}

\section{D-wave superconducting phase}

When there is only one short-range AFM correlation order parameters $\Delta
_{0}\neq 0$, the MF model Hamiltonian is simplified by%
\begin{align}
H{_{mf}^{\prime }}& =\sum_{\mathbf{k}\sigma }\varepsilon _{\mathbf{k}}c_{%
\mathbf{k}\sigma }^{\dag }c_{\mathbf{k}\sigma }-\frac{J_{K}V}{2}\sum_{%
\mathbf{k}\sigma }\left( f_{\mathbf{k}\sigma }^{\dag }c_{\mathbf{k}\sigma
}+c_{\mathbf{k}\sigma }^{\dag }f_{\mathbf{k}\sigma }\right)  \notag \\
& \text{ \ \ }+\lambda \sum_{\mathbf{k}\sigma }f_{\mathbf{k}\sigma }^{\dag
}f_{\mathbf{k}\sigma }+J_{H}\sum_{\mathbf{k}}\Delta (\mathbf{k})(f_{\mathbf{k%
}{\uparrow }}^{\dag }f_{-\mathbf{k}\downarrow }^{\dag }+h.c.)  \notag \\
& \text{ \ \ }+N\left( \frac{J_{K}V^{2}}{2}+J_{H}\Delta _{0}^{2}-\lambda
\right) ,
\end{align}%
where the spinons have a flat band. Due to the hybridization effect, the
heavy quasiparticle bands form. On the other hand, the spinon singlet
pairings imply a special non-magnetic state with short-range AFM
correlations among the local magnetic moments. Although there are no any
direct attractions among the conduction electrons, the spinon singlet
pairings provide an indirect glue for the formation of the Cooper pairs via
the Kondo screening/hybridizing effect. So the resulting MF ground state
represents a heavy fermion superconducting state.

On a two-dimensional square lattice, the spinon singlet pairing function has
a d-wave symmetry\cite{Kotliar-Liu}. Whether there are nodes or not mainly
depends on the Lagrangian multiplier $\lambda $ and hybridization strength $%
V $. From the simplified quasiparticle bands
\begin{eqnarray}
E_{\mathbf{k}}^{\pm } &=&\sqrt{E_{\mathbf{k},1}\pm \sqrt{E_{\mathbf{k}%
,1}^{2}-E_{\mathbf{k},2}^{2}}},  \notag \\
E_{\mathbf{k}1} &\equiv &\frac{1}{2}\left( \varepsilon _{\mathbf{k}%
}^{2}+\lambda ^{2}+J_{H}^{2}\Delta _{\mathbf{k}}^{2}\right) +\frac{1}{4}%
(J_{K}V)^{2},  \notag \\
E_{\mathbf{k}2} &\equiv &\sqrt{(\varepsilon _{\mathbf{k}}\lambda -\frac{%
J_{K}^{2}V^{2}}{4})^{2}+(\varepsilon _{\mathbf{k}}J_{H}\Delta _{\mathbf{k}%
})^{2}},
\end{eqnarray}%
we note that nodes appear in the lower band $E_{\mathbf{k}}^{-}$ when $E_{%
\mathbf{k},2}=0$, which requires the condition $\varepsilon _{\mathbf{k}}=%
\frac{J_{K}^{2}V^{2}}{4\lambda }$ in the diagonal direction.

Then the numerical mean field calculations are carefully performed with the
parameters $t^{\prime }/t=0.3$, $n_{c}=0.8$ and $J_{K}/t=2.0$. The lower
branch of the quasiparticle band $E_{\mathbf{k}}^{-}$ in the first Brillouin
zone along the momentum direction ($0,0$)$\rightarrow $($\pi /2,\pi /2$)$%
\rightarrow $($\pi ,\pi $)$\rightarrow $($\pi ,\pi /2$)$\rightarrow $($\pi
,0 $) are plotted in Fig.2 for different local AFM coupling strength $%
J_{H}/t=0.5$, $0.8$, $1.192$, and $1.3$, respectively. We first notice that
the quasiparticle states near the Fermi energy have a very small dispersion,
reflecting the quasiparticle mass enhancement in the superconducting state.
In Fig.2a and Fig.2b, a node can be clearly seen between the momentum ($\pi
/2,\pi /2$) and ($\pi ,\pi $) in the diagonal direction. Secondly, the
position of the node changes as increasing the local AFM coupling strength.
When the nodal position is shifted to ($\pi ,\pi $), further increasing $%
J_{H}/t$ makes a small energy gap open. The critical value can be determined
as $J_{H}/t=$ $1.192$. When $1.192<J_{H}/t<1.34$, the resulting
superconducting state has a full superconducting gap.

\begin{figure}[tbp]
\includegraphics[scale=0.4]{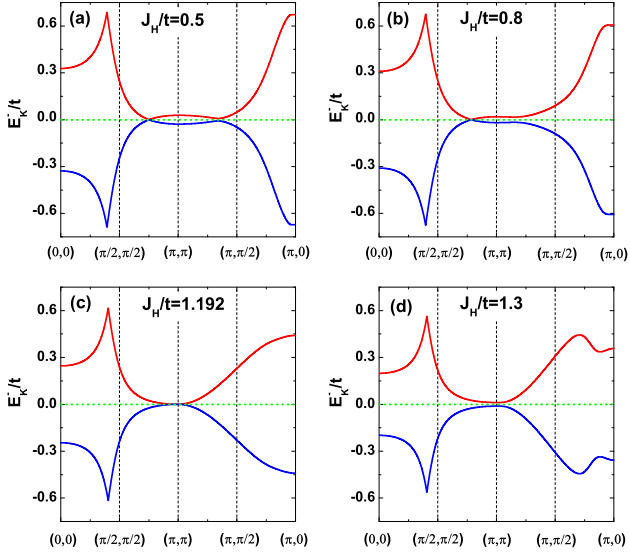}
\caption{(Color online) The lower branch of the superconducting
quasiparticle band struture for different local AFM Heisenberg coupling
strength with $t^{\prime }/t=0.3$, $n_{c}=0.8$ and $J_{K}/t=2.0$.}
\end{figure}

Moreover, the Kondo screening parameter $V$ and the AFM spinon pairing
parameter $\Delta _{0}$ have also been calculated, and the numerical results
are shown in Fig.3. It can be seen that, as $J_{H}/t$ starts to increase,
the Kondo screening parameter is almost unchanged in the beginning and then
becomes gradually decreased when $J_{H}/t>0.80$. On the contrary, the spinon
pairing order parameter $\Delta _{0}$ is extremely small in the small limit
of $J_{H}/t$. Only after $J_{H}/t>0.20$, $\Delta _{0}$ starts to grow up
quickly. Actually, we have also plotted a superconducting order parameter $%
\Delta _{sc}$ defined by the conducting electrons, whose definition will be
given later. It should be pointed out that the quantum phase transition
mentioned above does not manifest in the MF order parameters.

\begin{figure}[tbp]
\includegraphics[scale=0.3]{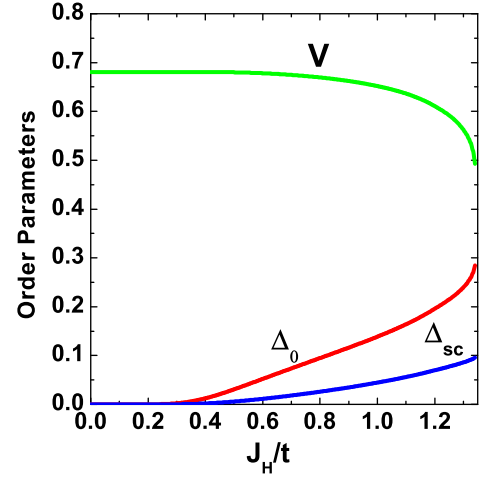}
\caption{(Color online) The effective mean field order parameters of the
Kondo screening $V$, the f-fermion/spinon pairing $\Delta _{0}$, and the
Cooper pairing $\Delta _{sc}$ for a given value of $J_{K}/t=2.0$.}
\end{figure}

As the spinons are paired up in the real lattice space, it is more
interesting to display the spinon pairing distribution in the momentum
space. Using the method of the equation motion, we can easily derive the
double-time retarded Green function for the spinon pairs,%
\begin{equation}
\langle \langle f_{-\mathbf{k}\downarrow }^{\dag }|f_{\mathbf{k}\uparrow
}^{\dag }\rangle \rangle _{\omega +i0^{+}}=\frac{J_{H}\Delta _{\mathbf{k}%
}(\omega ^{2}-\varepsilon _{k}^{2})}{\left[ \omega ^{2}-\left( E_{\mathbf{k}%
}^{+}\right) ^{2}\right] \left[ \omega ^{2}-\left( E_{\mathbf{k}}^{-}\right)
^{2}\right] },  \label{f1f2}
\end{equation}%
from which the expectation value of the spinon pairing function can be
evaluated according to the spectral theorem of the Green's function. At zero
temperature, we obtain%
\begin{equation}
\left\langle f_{\mathbf{k}\uparrow }^{\dag }f_{-\mathbf{k}\downarrow }^{\dag
}\right\rangle =-\frac{J_{H}\Delta _{\mathbf{k}}}{2\sqrt{2\left( E_{\mathbf{k%
}1}+E_{\mathbf{k}2}\right) }}\left[ 1+\frac{\varepsilon _{k}^{2}}{E_{\mathbf{%
k}2}}\right] ,
\end{equation}%
which is displayed in Fig.4. When the local AFM coupling strength $%
J_{H}/t=0.5$, we observe in Fig.4a that the spinon pairing function reflects
the d-wave symmetry around the Fermi surface of heavy quasiparticles, which
is a hole-like circle around the corner of the first Brillouin zone. A node
is clearly seen between the momentum ($\pi /2,\pi /2$) and ($\pi ,\pi $) in
the diagonal direction. The positive pairing magnitude and the negative
pairing magnitude are separated by the nodes. As increasing of $J_{H}/t$,
the spinon pairing region is outstretched and the position of the node is
shifted toward to ($\pi ,\pi $), as shown in Fig.4b. When $J_{H}/t=1.192$,
the node starts to disappear, and a small energy gap opens at ($\pi ,\pi $).
\begin{figure}[tbp]
\includegraphics[scale=0.4]{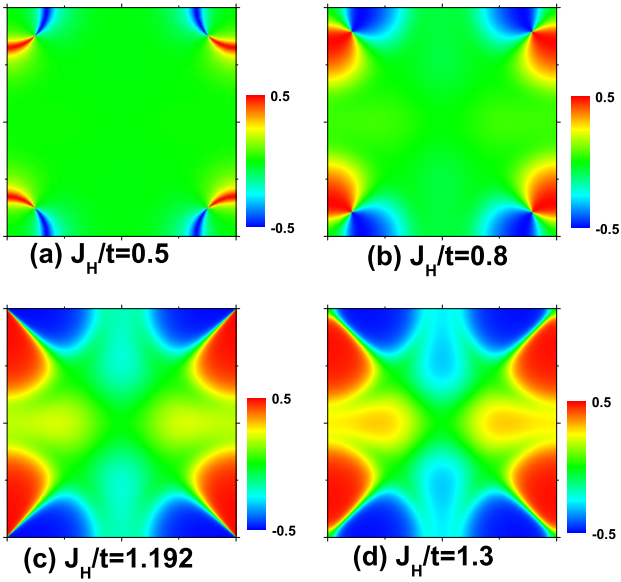}
\caption{(Color online) The spinon singlet pairing distribution in the first
Brillouin zone as increasing local AFM coupling strength for a given value
of $J_{K}/t=2.0$..}
\end{figure}

However, the Cooper pairing function of the conduction electrons is more
important for the heavy fermion superconducting state. Similarly, we can
easily derive the double-time retarded Green function for the conduction
electron pairs%
\begin{equation}
\langle \langle c_{-\mathbf{k}\downarrow }^{\dag }|c_{\mathbf{k}\uparrow
}^{\dag }\rangle \rangle _{\omega +i0^{+}}=\frac{J_{H}J_{K}^{2}V^{2}\Delta _{%
\mathbf{k}}}{4\left[ \omega ^{2}-\left( E_{\mathbf{k}}^{+}\right) ^{2}\right]
\left[ \omega ^{2}-\left( E_{\mathbf{k}}^{-}\right) ^{2}\right] },
\label{c1c2}
\end{equation}%
and the superconducting pairing function can be defined from the expectation
value of the conduction electron pairs. At zero temperature, we have
\begin{equation}
\left\langle c_{\mathbf{k}\uparrow }^{\dag }c_{-\mathbf{k}\downarrow }^{\dag
}\right\rangle =\frac{J_{H}J_{K}^{2}V^{2}\Delta _{\mathbf{k}}}{8E_{\mathbf{k}%
2}\sqrt{2\left( E_{\mathbf{k}1}+E_{\mathbf{k}2}\right) }},
\end{equation}%
which has been displayed in Fig.5. When $J_{H}/t=0.5$, we can clearly
observe that the positive pairing is distributed around the upper part of
the Fermi surface, while the negative pairing is around the lower part of
the Fermi surface. Nodes sit in the diagonal directions of the Fermi
surface. In contrast to the spinon pairing distribution, there are
additional features: a positive electron pairing magnitude is centered
around the point ($0,\pi $) and a negative electron pairing magnitude is
around the point ($\pi ,0$). As the local AFM coupling $J_{H}/t\geq 1.192$,
an energy gap opens near the nodes of the Fermi surface, and there are only
small pairing distribution on other parts of the Fermi surface. However, the
Cooper pairing distributed around momentum ($0,\pi $) and ($\pi ,0$) have
strong magnitudes. The resulting superconductivity still exhibits the d-wave
symmetry, i.e., the d-wave nodeless superconducting state.
\begin{figure}[tbp]
\center \includegraphics[scale=0.4]{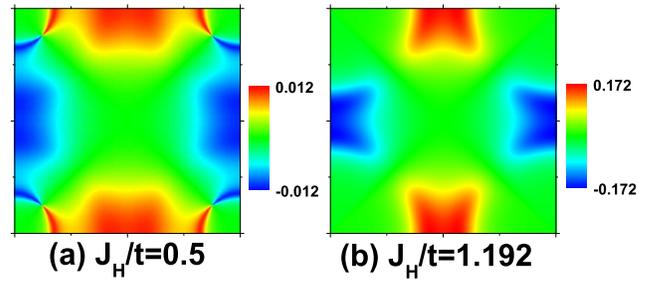}
\caption{(Color online) The superconducting pairing distribution function in
the first Brillouin zone for the local AFM coupling strength $J_{H}/t=0.5$
and $1.192$ with a given value of $J_{K}/t=2.0$.}
\end{figure}

Moreover, a real superconducting order parameter can be defined by
\begin{equation}
\Delta _{sc}=\frac{1}{2N}\sum_{\mathbf{k}}\left( \cos k_{x}-\cos
k_{y}\right) \langle c_{\mathbf{k}\uparrow }^{\dag }c_{-\mathbf{k}\downarrow
}^{\dag }+c_{-\mathbf{k}\downarrow }c_{\mathbf{k}\uparrow }\rangle ,
\end{equation}%
which has been displayed in Fig.3. Since it is generated by the spinon
singlet pairings, the superconducting order parameter has a smaller
magnitude compared to the spinon pairing order parameter $\Delta _{0}$. In
order to reveal the quantum phase transition between two superconducting
phases, we have to calculate the first order derivative of the ground state
energy. For a given value of $J_{K}/t=2.0$, the numerical results are
exhibited in Fig.6. At $J_{H}/t=1.192$, the ground state energy and its
first derivative are continuous but the second derivative is not. This
corresponds the critical point of a second order quantum phase transition.
\begin{figure}[tbp]
\center \includegraphics[scale=0.4]{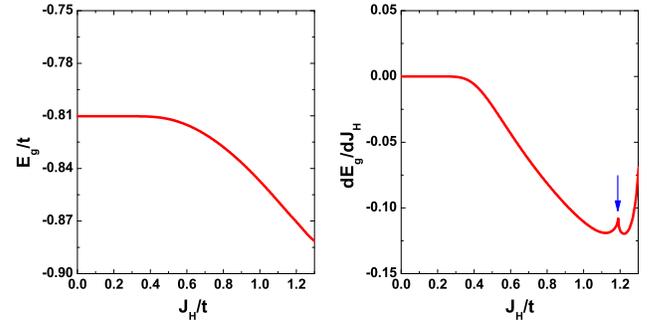}
\caption{(Color online) The ground state energy density (a) and its first
order derivative (b) as a function of $J_{H}/t$ for a given value of $%
J_{K}/t=2$. The arrow indicates the critical point of quantum phase
transition.}
\end{figure}

\section{Discussion and conclusion}

So far we have presented the ground state properties of the effective MF
theory for the two-dimensional Kondo lattice model with the local AFM
Heisenberg exchange coupling between the localized magnetic moments. We
would like to emphasize that, it is the local AFM short-range interaction
that induces the AFM and superconducting long-range ordering states. When $%
J_{K}\gg J_{H}$, the local AFM Heisenberg exchange coupling among the local
moments is negligible and the stable paramagnetic heavy Fermi liquid state
is resulted from the coherent superposition of individual Kondo screening
clouds. When $J_{H}$ is increased but still less than $J_{K}$, the local AFM
Heisenberg exchange coupling has to be taken into account, and the spinon
singlet pairings can reduce the ground state energy significantly. The
Cooper singlet pairs among the conduction electrons can be induced via the
Kondo screening effect, leading to the heavy fermion superconductivity. This
also belongs to the category of the unconventional superconductivity
mediated by the short-range AFM correlations. However, in the present MF
theory, we do not find a finite critical value of $J_{H}/t$ to separate the
heavy fermion liquid and the superconducting phases. This may be caused by
the approximation used in the effective MF theory.

The resulting superconducting state exhibits the d-wave symmetry. Whether
the Cooper pairing function has nodes or not depends on the ratio of the AFM
Heisenberg and Kondo exchange couplings. As the coupling ratio increases,
the nodal position is shifted outward along the direction of ($\pi /2,\pi /2$%
)$\rightarrow $($\pi ,\pi $). When the nodal position reaches the end point
of the Brillouin zone, a quantum phase transition occurs, and a full
superconducting gap is opened at the Fermi energy. As further increasing of $%
J_{H}/J_{K}$, the present MF theory is no longer valid, because the
long-range AFM correlations have to be taken into account.

In conclusion, a possible mechanism of heavy fermion superconductivity with
d-wave symmetry is carefully investigated in the two-dimensional
Kondo-Heisenberg lattice model from the limit of the paramagnetic heavy
Fermi liquid $J_{K}\gg J_{H}$. The resulting d-wave superconducting
properties can be related to some heavy fermion superconductors with the
similar structure like CeCoIn$_{5}$, where thermal conductivity measurements
strongly support a superconducting gap with nodes along the diagonal
directions in the Brillouin zone\cite{Izawa}. Further theoretical work
including the estimation of the fluctuations around the MF solution or the
coexistent phase of both the AFM order and superconductivity would be
necessary to be considered.

The authors would like to thank T. Xiang for stimulating discussions and
acknowledge the support of NSF of China and the National Program for Basic
Research of MOST-China.

\end{document}